\documentstyle[aps,manuscript]{revtex}  
\makeatletter
\newcounter{subeqncnt}
\def\thesubeqncnt{\alph{subeqncnt}}
\def\subequations{\begingroup%
   \stepcounter{equation}\edef\@tempa{\theequation}%
   \let\c@equation\c@subeqncnt\c@subeqncnt\z@
   \edef\theequation{\@tempa\noexpand\thesubeqncnt}}

\makeatother

\begin{document}                
\title{Convergent recursive O($N$) calculations for
       $ab~initio$ tight-binding}
\author{T. Ozaki$^{1,2}$ and K. Terakura$^{1,3}$}
\address{
     $^{(1)}$ RICS,
     National Institute of Advanced Industrial Science and Technology (AIST),
     central 2, 1-1-1 Umezono, Tsukuba,
     Ibaraki 305-8568, Japan
}
\address{
     $^{(2)}$ JRCAT-ATP,
     central 4, 1-1-1 Higashi, Tsukuba, 
     Ibaraki 305-0046, Japan
}
\address{
     $^{(3)}$ JRCAT-AIST,
     central 4, 1-1-1 Higashi, Tsukuba, 
     Ibaraki 305-8562, Japan
}
\maketitle
\begin{abstract}                
     A theory is presented for a novel recursion method for O($N$)
     $ab~initio$ tight-binding calculations.
     A long-standing problem of generalizing the recursion method to
     a non-orthogonal basis, which is a crucial step to make the recursion
     method applicable to $ab~initio$ calculations, is solved by the 
     introduction of the hybrid representation and the two-sided block
     Lanczos algorithm. 
     As a test of efficiency of the new method, convergence properties
     in energy and force of insulators, semiconductors, metals, and molecules
     are studied for not only simple model systems but also some real
     materials within the density functional theory.
     The present O($N$) method possesses good
     convergence properties for metals as well as insulators.
\end{abstract}
\vspace{2cm}

 The application of the conventional $ab~initio$ electronic
 structure calculations to large systems is hampered by their inherent
 O($N^3$) scaling properties with $N$
 the number of atoms. In order to overcome this difficulty,
 efficient linear scaling algorithms,
 which are referred to as O($N$) methods, have been developed during the
 last decade \cite{Pettifor,Ozaki,ON}.
 A few applications of these O($N$) methods have actually demonstrated
 the power of these O($N$) methods \cite{Applications}.
 However, several problems still remain in these O($N$) methods.
 In particular, it is well known that the variational O($N$) methods
 produce large errors in the energy of metals \cite{Comparison}, since in
 both metals and narrow-gap semiconductors the density
 matrix $\rho$ has long range correlations in real space compared
 to that of insulators with a wide gap \cite{Arias}.
 Therefore, the application of the O($N$) methods relying on the locality
 of $\rho$ is restricted to materials with a wide gap.

 Recently, one of the authors has shown that the block bond-order
 potential (BOP) is a convergent moment-based O($N$) method which
 provides good convergence properties for energy and force
 in metals as well as insulators and semiconductors within an
 orthogonal tight-binding (TB) representation \cite{Ozaki}.
 Although it had been well documented that the moment-based
 methods cannot reproduce the vacancy in diamond or silicon
 within a low number of moments \cite{Comparison,Voter},
 the block BOP gave the first convergent results for vacancies
 in insulators using a moment-based method with a low
 number of moments \cite{Ozaki}.
 The block algorithm guarantees that the $\sigma$ and $\pi$ bonds
 are treated properly
 and that the band energy is invariant for the rotation of
 systems \cite{Jones}, and the terminator in the Green functions
 can estimate the long range contributions in a most
 effective way \cite{Haydock} for the total energy and force
 calculations. Thus, the block BOP may be applicable
 to a wide variety of materials with reliable accuracy.
 However, in order to make the method applicable to $ab~initio$
 calculations, we have to remove the limitation of orthogonal basis
 and reformulate the method in the non-orthogonal basis.
 Several attempts have been made to generalize the recursion method
 to non-orthogonal basis~\cite{Haydock}.
 Nevertheless, within our knowledge the formalism is not
 in a satisfactory level and only very few examples of
 $ab~initio$ calculations have been available~\cite{Gibson}.
 In this Letter we present a new generalized
 block BOP for the non-orthogonal TB basis by introducing
 a hybrid representation and the two-sided block Lanczos algorithm.
 We will show convergence properties of the method within density
 functional theory (DFT) and demonstrate that the present method
 is a promising and practical O($N$) method in both insulators and
 metals.

 It will be assumed that one-particle wave functions are expanded
 with a non-orthogonal localized basis set $(\vert i\alpha\rangle)$
 where $i$ is a site index, and $\alpha$ an orbital index.
 Then, the density of electrons $\rho({\bf r})$ in non-spin polarized
 systems can be written as
 \begin{eqnarray}
   \rho({\bf r}) = \sum_{i\alpha,j\beta}
                   \chi_{i\alpha}({\bf r})\chi_{j\beta}({\bf r})
                   \Theta_{i\alpha,j\beta},
 \end{eqnarray}
 where $\chi_{i\alpha}({\bf r})\equiv \langle{\bf r}|i\alpha\rangle$,
 and the bond-order $\Theta_{i\alpha,j\beta}$ is related to
 the imaginary part of the one particle Green function as follows:
 \begin{eqnarray}
    \Theta_{i\alpha,j\beta} = -\frac{2}{\pi}
                {\rm Im}\int G_{i\alpha,j\beta}(E+{\rm i0^+})
                f(\frac{E-\mu}{k_BT})dE
 \end{eqnarray}
 with the Fermi function $f(x)=1/[1+{\rm exp}(x)]$,
 where $0^+$ represents a positive infinitesimal.
 $G_{i\alpha,j\beta}(Z)$ is a expectation value of
 Greenian $\hat{G}(Z)\equiv (Z-\hat{H})^{-1}$ for dual bases
 $\vert \tilde{i\alpha}\rangle$ and $\vert \tilde{j\beta}\rangle$
 defined by
 \begin{eqnarray}
    |\tilde{i\alpha} \rangle = \sum_{j\beta}|j\beta\rangle
                               S_{j\beta,i\alpha}^{-1},
 \end{eqnarray}
 where $S_{j\beta,i\alpha}^{-1}$ is the inverse of overlap matrix
 $S_{i\alpha,j\beta}\equiv \langle i\alpha \vert j\beta \rangle$.
 From Eqs.~(1) and (2), we see that functionals such as the total energy
 in DFT can be reformulated as the functionals of the bond-order.
 Therefore, we concentrate on evaluating the bond-order with
 reasonable accuracy and great reduction in the computational efforts.

 Let us introduce a hybrid representation of Hamiltonian which
 is a non Hermitian matrix represented by the original and
 the dual bases as
 $  H'_{i\alpha,j\beta} =
    \langle \tilde{i\alpha}\vert \hat{H}\vert j\beta \rangle. $
 The hybrid Hamiltonian can be written in the matrix form as
 $H'=S^{-1}H$,
 where $H_{i\alpha,j\beta}\equiv
 \langle i\alpha\vert \hat{H} \vert j\beta \rangle$.
 With the relation $G(Z)(ZS-H)={\rm I}$,
 the hybrid Green function $G'(Z)$ defined by
 \begin{eqnarray}
    G'_{i\alpha,j\beta}(Z) = \{G(Z)S\}_{i\alpha,j\beta}
     = \langle \tilde{i\alpha}\vert \hat{G}(Z)\vert j\beta \rangle
 \end{eqnarray}
 satisfies $G'(Z)(Z{\rm I}-H')={\rm I}$.
 One of the merits of using $G'(Z)$ is
 that its diagonal element gives
 directly the Mulliken population $P_{i\alpha}$ of an orbital
 $\vert i\alpha\rangle$:
 \begin{eqnarray}
    \nonumber
    P_{i\alpha} & = & -\frac{2}{\pi}{\rm Im}\int
                      G'_{i\alpha,i\alpha}(E+0^+)
                      f(\frac{E-\mu}{k_BT})dE \\
                & = &
                  \sum_{j\beta}\Theta_{i\alpha,j\beta}S_{j\beta,i\alpha}.
 \end{eqnarray}
 In the block BOP, determination of the chemical potential is needed
 to conserve the total number of electrons $N_{ele}$ in the
 system \cite{Ozaki}, so that the relation of Eq.~(5) is very
 advantageous to computational efficiency because of the simple
 relation $N_{ele}=\sum_{i\alpha}P_{i\alpha}$.
 Thus, we present below a prescription how to calculate the hybrid
 Green functions.
 The diagonal elements of the Green function matrix can be calculated
 in a numerically stable way by the recursion method \cite{Haydock}
 based on the Lanczos algorithm \cite{Lanczos}.
 The block BOP method is a general recursion method for evaluating
 efficiently both the diagonal and off-diagonal elements of the
 Green function matrix by the recursion method.
 Moreover the use of a single site containing all the localized
 orbitals as the starting state in the {\it block} Lanczos algorithm
 rather than a single orbital in the usual one conserves
 the rotational invariance of the total energy.
 In the present case of non-orthogonal basis, we further
 extend the formalism to adopt a two-sided
 block Lanczos algorithm \cite{TSL}, since the hybrid Hamiltonian is
 not any more Hermitian.
 The set of central equations are
 \begin{eqnarray}
   \nonumber
   \hat{H}\vert U_{n}) & = & \vert U_{n})\underline{A}_{n}
                       +
            \vert U_{n-1})\underline{B}_{n}
                       +
            \vert U_{n+1})\underline{C}_{n+1},\\[2mm]
   (\tilde{U}_{n}\vert \hat{H} & = & \underline{A}_{n}(\tilde{U}_{n}\vert
                       +
                      \underline{C}_{n}(\tilde{U}_{n-1}\vert
                       +
                      \underline{B}_{n+1}(\tilde{U}_{n+1}\vert.
 \end{eqnarray}
 $\underline{A}_n$, $\underline{B}_n$, and $\underline{C}_n$
 are recursion block coefficients with $M_i\times M_i$ in size,
 where $M_i$ is the number of localized orbitals on the
 starting atom $i$, and the underline indicates that the element is a block.
 In the two-sided block Lanczos algorithm the Lanczos vectors in the
 left and right sides have a bi-orthogonality relation.
 It is essential to start the two-sided block Lanczos algorithm with
 a single site and its corresponding dual state as
 \begin{eqnarray}
    \nonumber
    \vert U_0) & = &
            (\vert i1\rangle,\vert i2\rangle,\dots,\vert iM_i\rangle ),\\[2mm]
    \vert \tilde{U}_0) & = &
            (\vert \tilde{i1}\rangle,\vert \tilde{i2}\rangle,\dots,
             \vert \tilde{iM_i}\rangle ).
 \end{eqnarray}
 Equation~(7) is an optimum choice in terms of computational accuracy
 and efficiency because of the rotational invariance of the total energy
 and the consistent description for the different properties of
 $\sigma$, $\pi$, and $\delta$ bonds.

 In the Lanczos basis representation the Hamiltonian $H^L$ is
 block-tridiagonalized as a non Hermitian matrix and the Green
 function matrix $G^L(Z)$ is the inverse of the matrix
 $(Z{\rm I}-H^{L})$, so that the block diagonal element
 $\underline{G}^{L}_{00}(Z)=(\tilde{U}_{0}\vert \hat{G}\vert U_{0})$
 can be written explicitly in the form of the multiple inverse as follows:
 \begin{eqnarray}
   \nonumber
   \underline{G}^L_{00}(Z)
        =[Z\underline{\rm I}-\underline{A}_0-\underline{C}_1[
          Z\underline{\rm I}-\underline{A}_1-\underline{C}_2[
                       \cdots
          ]^{-1}\underline{B}_2
          ]^{-1}\underline{B}_1
          ]^{-1},\\
 \end{eqnarray}
 where the index $L$ indicates the representation based on the Lanczos basis.
 The off-diagonal elements of hybrid Green function matrix can be
 calculated by using a recurrence relation which can be derived
 basically along the same line as that described for the
 case of orthogonal basis \cite{Ozaki}.  The explicit expression
 consistent with Eqs. (6) and (8) is given below:
 \begin{eqnarray}
    \nonumber
    \lefteqn{
      \underline{G}^{L}_{0n}(Z)
      =
    \biggl(
      \underline{G}^{L}_{0n-1}(Z)(Z\underline{\rm I}
       -\underline{A}_{n-1})
    }\\
    &&
    \quad\quad\quad\quad
       -\underline{G}^{L}_{0n-2}(Z)
         \underline{C}_{n-1}
         -\delta_{1n}\underline{\rm I}
    \biggr)
         (\underline{B}_{n})^{-1},
 \end{eqnarray}
 where $\delta$ is Kronecker's delta, and $\underline{G}_{0-1}(Z)=
 \underline{C}_{0}= \underline{0}$.
 The block elements of the Green function matrix
 have the same relation to the bond-orders based on the Lanczos basis
 $\underline{\Theta}^L_{0n}$ as
 Eq.~(2) of the dual basis representation. Therefore,
 we can obtain the bond orders of Eq.~(2) through the following
 transformation:
 \begin{eqnarray}
    \underline{\Theta}_{ij} = \sum_{n,k}
                         \underline{\Theta}^L_{0n}
                         \underline{\tilde{U}}_{nk}\underline{S}_{kj}^{-1},
 \end{eqnarray}
 where $\underline{\tilde{U}}_{nj}$ is defined by
 $ \underline{\tilde{U}}_{nj} = (\tilde{U}_{n}\vert
        (\vert j1\rangle,\vert j2\rangle,\dots,\vert jM_j\rangle ).$
 As a result of the simple inverse transformation Eq.~(10),
 we only have to perform the evaluation and the integration
 of the Green functions of the 0th block line in the Lanczos basis
 representation, which means that the computational time of
 the algorithm is about two times \cite{com0} longer compared
 to that of the orthogonal case \cite{Ozaki}.
 Only the hybrid representation can provide this simple relation Eq.~(10)
 as well as Eq.~(5), while the other representations suffer from
 computational inefficiency \cite{Haydock,Gibson}.
 In the generalized block BOP using the non-orthogonal basis we need to
 calculate $S^{-1}$, the inverse of the overlap matrix.
 In the following calculations, we used a new O($N$) efficient method
 for inverting the overlap matrix \cite{com1}.

 In Fig.~1 we show convergence properties of the band energy
 in an insulator and a metal described by a simple s-valent TB
 as a test of the present method. The errors in the band energy
 at the seven-shell cluster and recursion levels are 0.2~\% and 0.9~\%
 for the insulator and the metal, respectively.
 Thus, we see that the block BOP gives sufficient convergent results
 in both the simple insulator and metal.
 Figures 2(a) and (b) show the error in the band energy at
 the five-shell cluster and recursion levels for insulators and metals
 described by a simple s-valent TB as a function of direct band gap
 and electronic temperature, respectively.
 In insulators the error goes to zero as the gap increases,
 while the errors, whose absolute values are no more than 0.5~\%
 compared to the band energy in the whole region, are relatively small.
 In metals the error becomes almost negligible for the higher
 electronic
 temperature. This behavior in both insulators and metals
 is consistent with the recent study about the locality of the density
 matrix \cite{Arias}, though the block BOP depends on the convergence
 of the moment expansions for the density matrix rather than
 the locality of the density matrix \cite{Ozaki}.
 From the comparison in the NaCl and FCC structures it is clear that
 the use of the terminator in the diagonal Green functions effectively
 reduces the error in both cases.

 Next we discuss convergence properties of the block BOP in realistic
 materials within the TB based DFT proposed by Sankey and
 Niklewski \cite{Sankey}. Figure 3 shows the convergence properties of
 the cohesive energy for carbon in the diamond structure, silicon in the
 diamond structure, fcc aluminum, and C$_{60}$ molecule \cite{com2}.
 In carbon and silicon the cohesive energies rapidly converge
 to the k-space results in the five and seven-shell clusters,
 while the convergence values for the three-shell cluster are
 in error by 0.4 and 0.9~\% from the k-space results, respectively.
 Even for metallic aluminum, the convergence is very fast 
 with respect to the number of recursion levels and the errors 
 in the converged values are only 0.3 and 0.1~\% for the
 three- and five shell clusters, respectively.
 For C$_{60}$ the convergence is achieved with the three-shell cluster.
 The error at the sixth recursion level is only 0.02~\%.
 As a test of the consistency between the total energy and the forces,
 a constant energy molecular dynamics simulations have been performed
 for diamond within DFT.
 In Fig.~4 we show the energy for diamond at 300~K using five and
 ten recursion levels.
 For five and ten recursion levels we see that the total energy
 is almost conserved, while the ten recursion level calculation gives a better
 result.  These test calculations indicate that the block
 BOP can give forces consistent
 with the total energy.

 In summary, we have presented a theory of the block BOP for O($N$)
 $ab~initio$ TB calculations. The introduction of the hybrid representation
 and the two-sided block Lanczos algorithm enables us to generalize the
 theory to a non-orthogonal basis set in a natural way.
 The test calculations for the simple s-valent TB systems and some
 real systems within DFT
 suggest that the O($N$) method provides a rapid convergence
 properties for metals as well as insulators with sufficient accuracy.
 Thus, we conclude that the block BOP is a robust
 O($N$) methods which is applicable to a wide variety of materials
 in the $ab~initio$ TB approach.

 We would like to thank Y. Morikawa and H. Kino for helpful suggestions
 about the DFT calculations.

%

 
 \begin{figure}[t]
  \caption{\small
   The error, with respect to the standard k-space calculations,
   in the band energy for an insulator (zinc blende) and a metal
   (FCC) described by a simple s-valent TB model in which the
   nearest neighbor hopping and overlap integrals are -1.0~eV and
   0.1, respectively, with others being zero, and the number of
   electrons is the same as that of atoms.
   The zinc blende has a direct gap of 1.0~eV which was
   controlled by the gap of the on-site energies of
   the different atoms. In these calculations, the seven-shell
   cluster and a square-root terminator were used.}
 \end{figure}


 \begin{figure}[t]
  \caption{\small
   The error in the band energy for (a) insulators and (b) metals,
   calculated at the five-shell cluster and recursion levels.
   The calculations were carried out with the same s-valent
   TB model as that in Fig.~1 using a square-root terminator.
   For NaCl and FCC the non terminator results are also shown.
 }
 \end{figure}


 \begin{figure}[t]
  \caption{\small
   The error in the cohesive energy for carbon in the diamond
   structure, silicon in the diamond structure, fcc aluminum, 
   and C$_{60}$ for three-, five-, and seven-shell clusters,
   calculated using a square-root terminator.
   These calculations were performed within DFT.}
 \end{figure}


 \begin{figure}[t]
  \caption{\small
   The potential, kinetic, and total energies as a function of 
   time for molecular dynamics simulations of carbon using 
   a three shell cluster and a square root terminator. 
   In panels (a) and (b) the results are for five and ten recursion
   levels at 300~K, respectively. The time step is 0.5~fs.}
 \end{figure}


\begin{references}
\bibitem{com0} For diamond, the computational time on a sharp
  PC-MJ140M taken per MD step per atom are 1.8 and 2.8 sec for
  the orthogonal and non-orthogonal cases, respectively.
\bibitem{com2}
   In these density functional calculations minimal basis sets for
   valence electrons were used for carbon, silicon, and aluminum atoms.
   The radii of the radial-wave function confinement are 2.1, 3.1, and
   3.7~\AA, respectively.
   The minimal basis sets give 1.253 (1.244), 2.280 (2.246),
   and 2.515 (2.466) \AA~as an equilibrium bond length of dimer for carbon,
   and silicon, and aluminum, respectively, where the values in the
   parentheses are experimental results.
\end{references}
\end{document}